\documentclass[11pt,aps,prd,preprint,preprintnumbers,showpacs,superscriptaddress,floatfix]{revtex4-1}
\usepackage{graphicx}
\usepackage{epsfig}
\usepackage{dcolumn}
\usepackage{bm}
\usepackage{subfigure}
\usepackage{amsmath}
\usepackage{amssymb}
\usepackage{bbm}
\usepackage{multirow}
\usepackage{setspace}
\usepackage{xcolor}
\usepackage[colorlinks,
            linkbordercolor={0 0 0},
            pdfborder={0 0 0},
            linkcolor=blue,
            citecolor=blue,
            urlcolor=blue,
            breaklinks=true]{hyperref}

\allowdisplaybreaks

\def\rmsmall #1{\mbox{\scriptsize #1}}

\def\eqarray#1{\begin{eqnarray} #1 \end{eqnarray}}
\newlength{\x}
\newlength{\y}
\newlength{\z}
\urldef\milcurl\url{http://www.physics.utah.edu/~detar/milc/}

\phantomsection
\begin{document}
\preprint{IMSc/2016/10/06}

\title{Confinement-Deconfinement transition in $SU(2)+$Higgs Theory}

\author{Minati Biswal}%
\email{mbiswal@imsc.res.in}%
\affiliation{The Institute of Mathematical Sciences, Chennai-600113 and
Homi Bhaba National Institute, Mumbai-400094, India}%

\author{Mridupawan Deka}%
\email{mpdeka@theor.jinr.ru}%
\affiliation{The Institute of Mathematical Sciences, Chennai-600113 and
Homi Bhaba National Institute, Mumbai-400094, India}%

\affiliation{Bogoliubov Laboratory of Theoretical Physics, JINR, 141980 Dubna, Russia}%

\author{Sanatan Digal}%
\email{digal@imsc.res.in}%
\affiliation{The Institute of Mathematical Sciences, Chennai-600113 and
Homi Bhaba National Institute, Mumbai-400094, India}%

\author{P. S. Saumia}%
\email{saumia@theor.jinr.ru}%
\affiliation{The Institute of Mathematical Sciences, Chennai-600113 and 
Homi Bhaba National Institute, Mumbai-400094, India}%

\affiliation{Bogoliubov Laboratory of Theoretical Physics, JINR, 141980 Dubna, Russia}%

\begin{abstract}
  
We study the confinement-deconfinement transition in $SU(2)$ gauge theory in the
presence of massless bosons using lattice Monte Carlo simulations. The nature of
this transition depends on the temporal extent ($N_\tau$) of the Euclidean 
lattice. We find that the transition is a cross-over for $N_\tau=2,4$ and second 
order with $3D$ Ising universality class for $N_\tau=8$. Our results show that
the second order transition is accompanied by realization of the $Z_2$ symmetry. 

\end{abstract}

\pacs{11.10.Wx,11.15.Ha,11.15.-q}

\maketitle

\section{Introduction} 
\label{sec:intro}  
 
Gauge theories such as quantum chromodynamics (QCD), standard model(SM) etc. at finite 
temperatures are relevant for describing the phase transitions in the early Universe 
and in the relativistic heavy-ion collisions. The pure gauge parts of these 
theories undergo the confinement-deconfinement (CD) transition~\cite{Kuti:1980gh,
McLerran:1980pk} at high temperatures. The corresponding pure gauge Euclidean actions are
invariant under a class of gauge transformations represented by the center
$Z_N$ of the $SU(N)$ group. This $Z_N$ symmetry~\cite{Svetitsky:1985ye,Svetitsky:1982gs} 
plays an important role in the CD transition. In many ways, the nature of the CD
transition is found to be similar to the transition in spin systems with $Z_N$ symmetry.
The $Z_N$ symmetry is spontaneously broken in the deconfined phase by a non-zero thermal 
expectation value of the Polyakov loop. This leads to $N$ degenerate phases in the 
deconfined state.

\smallskip

In the fundamental representation, the $Z_N$ symmetry is explicitly broken in the presence 
of the matter fields. The $Z_N$ group can act only on the gauge fields and its action on the 
matter fields spoils their necessary temporal boundary condition.
This explicit breaking affects the nature 
of the CD transition and the thermodynamic behavior of the phases themselves. It
weakens the CD transition and, in the deconfined phase, all but 
only one of the $N$ phases become meta-stable. The explicit breaking vanishes when 
the matter fields are infinitely heavy. So it is expected that the explicit $Z_N$ symmetry 
breaking is small for large dynamical masses of the matter fields. In the mean field 
approximation of QCD, the explicit symmetry breaking turns out to be an 
effective ``uniform" external field acting on the Polyakov loop~\cite{Green:1983sd}
when the fermion masses are large. The strength of the external field grows as the masses 
decrease. Non-perturbative studies find that the CD transition in $SU(2)$ gauge 
theory with dynamical fermions is a crossover~\cite{Nakamura:1984uz,
Heller:1984eq, Heller:1985wc, Kogut:1985xd, Kogut:1985un}. For $SU(3)$ gauge theory, 
the CD transition becomes a weak first order transition for large fermion 
mass~\cite{Polonyi:1984zt,Hasenfratz:1983ce,Gavai:1985nz,Fukugita:1986rr,Karsch:2001nf}. These
results are consistent with the findings of the mean field approximation. However, an
extrapolation of this effective external field to the chiral limit fails 
to explain the nature of the CD transition and the Polyakov loop behavior. In this case, 
the nature of the CD transition turns out to be the same as the chiral 
transition~\cite{Digal:2002wn,Fukushima:2002ew}. This suggests that, in the chiral limit, the 
effective external field is a  fluctuating and non-uniform dynamical field instead 
of a fixed uniform field.  The behaviour of the chiral transition and the chiral condensate
are, however, well described by a uniform/static field in the chiral limit~\cite{Pisarski:1983ms}.

\smallskip
It is expected
that the explicit breaking of $Z_N$ due to bosonic matter fields also depends on mass.
Perturbative calculations show that the explicit symmetry breaking increases with decrease
in mass in presence of fermionic matter fields~\cite{Gross:1980br,Weiss:1981ev}. A straightforward 
extension of these 1-loop calculations for bosonic fields gives similar results. 
For the massless case, the explicit symmetry breaking for 
$N=2$ is so large that there are no meta-stable states in the deconfined phase. These calculations, 
however, are not reliable near the CD transition. 
%Hence, non-perturbative lattice studies of 
%the effects of the matter fields near the CD transition are required. Studies of the lattice 
%non-abelian gauge theories coupled to the Higgs with fixed radial mode find that the 
%CD transition is stable
 Strong coupling studies of lattice non-abelian gauge theories coupled to the Higgs field with 
the fixed radial mode find that the CD transition behaves like a pure gauge CD transition even 
for some finite non-zero coupling between the gauge and Higgs fields~\cite{Fradkin:1978dv}. 
For heavy Higgs fields, non-perturbative calculations find that the temperature dependence of the 
Polyakov loop expectation value shows a critical behavior above the CD
transition point, i.e $\left<L\right> \sim (T-T_c)^{1\over 3}$~\cite{Damgaard:1986qe, 
Damgaard:1986jg}. Recent study of the $Z_N$ symmetry~\cite{Biswal:2015rul} 
shows, within the numerical errors, that the strength of the explicit symmetry breaking 
vanishes even for a large but finite Higgs mass. 
These results indicate clear deviations from those of perturbative calculations
in presence of matter fields. It is not clear whether the conventional expectation that the 
transition becomes weaker with the mass of matter fields, which is observed in QCD, also
holds in the case of $SU(N)+$Higgs. To address this issue, we study the CD transition in the
presence of the Higgs  with vanishing bare mass using non-perturbative Monte Carlo simulations. 
We also compare the non-perturbative and perturbative results away from CD transition. To 
simplify our study, we consider $N=2$ and vanishing Higgs quartic coupling.

\smallskip

From lattice simulations, it is known that the thermal average of the Polyakov
loop~\cite{Svetitsky:1985ye, Weiss:1981ev} has strong cut-off dependence. The Polyakov loop expectation 
value decreases with the number of temporal cites ($N_\tau$) of the Euclidean lattice. 
However, the nature of the pure gauge CD transition does not depend on
$N_\tau$~\cite{Datta:1999np,Kogut:1982rt}. 
In the presence of massless Higgs, this transition is found to 
be dependent on $N_\tau$. In this study, we find that this transition is a cross-over for 
$N_\tau=2,4$ and second order for $N_\tau=8$. These results suggest 
that in the continuum limit the CD transition is second order. We also look at the distribution 
of the Polyakov loop values in the thermal ensemble. The distribution in the case of $N_\tau=8$ clearly 
exhibits the $Z_2$ symmetry, which also explains why the CD transition is second 
order. This is surprising as one would expect maximal symmetry breaking as is
observed in perturbative calculations~\cite{Gross:1980br,Weiss:1981ev} as well as in lattice
QCD~\cite{Karsch:2001nf,Digal:2000ar}. Coincidentally the realization of the $Z_2(Z_N)$ symmetry 
occurs only when the system is in the Higgs symmetric phase. This suggests that the 
strength of the Higgs condensate may be playing the role of the effective external 
field for the CD transition.  We think that this restoration of the $Z_2(Z_N)$ symmetry
for larger $N_\tau$ is not due to the trivial continuum limit of pure Higgs 
theories~\cite{Callaway:1988ya}  since the interaction between the gauge and Higgs increases with 
$N_\tau$.  We discuss the possible reasons 
of this realization of $Z_2$ (or $Z_N$) symmetry in the Higgs symmetric phase 
later in section IV. 

\smallskip

The paper is organized as follows. In section II we describe the $Z_N$ symmetry in  
$SU(N)+$Higgs theory. In section III we describe our simulations and results for $N=2$. This 
is followed by conclusions in section IV.  
  
%%%%%%%%%%%%%%%%%%%%%%%%%%%%%%%%%%%%%%%%%%%%%%%%%%%%%%%%%%%%%%%%%%%%%%%%%%%%%%%%%%%%%%%%% 
\section{The $Z_N$ symmetry in the presence of fundamental Higgs fields}  
%%%%%%%%%%%%%%%%%%%%%%%%%%%%%%%%%%%%%%%%%%%%%%%%%%%%%%%%%%%%%%%%%%%%%%%%%%%%%%%%%%%%%%%%% 
 
The finite temperature partition function for a $SU(N)$ gauge field, $A_\mu$,
in the path-integral formulation is given by
\begin{equation} 
{\cal Z} = \int [DA] e^{-S_G}, 
\end{equation} 
\noindent with the following gauge action  
\begin{equation} 
\begin{aligned} 
S_G = \int_V d^3x\int_0^\beta d\tau {1\over 2}\left[  
Tr\left(F^{\mu\nu}F_{\mu\nu}\right)\right], 
F_{\mu\nu}=\partial_\mu A_\nu-\partial_\nu A_\mu + g[A_\mu,A_\nu].
\end{aligned} 
\end{equation} 
The gauge field for a given Euclidean component $\mu$ is a $N \times N$ matrix,  
$A_\mu=T^aA^a_\mu$, where $T^a$'s are the generators of the $SU(N)$ group. Here $\beta$ 
is the inverse of the temperature $T$. The  
path-integration is over all $A_\mu$'s which are periodic along the temporal  
direction $\tau$, i.e $A_\mu(\tau)=A_\mu(\tau+\beta)$. This periodicity allows 
the gauge transformations $U(\tau)$ to be non-periodic along the temporal direction, 
up to a factor $z\in Z_N$ as
\begin{equation} 
U(\tau=0)=z U(\tau=\beta). 
\label{bcu}
\end{equation} 
Though the action is invariant under such gauge transformations, the Polyakov loop
\begin{equation} 
L(\vec{\bf x})={1 \over N} {\rm{Tr}}\left[{\rm{P}}\left\{\exp{\left(-ig\int_0^\beta  
A_0 d\tau\right)}\right\}\right], 
\end{equation} 
transforms as $L \longrightarrow z L$. In the deconfined phase $L$ acquires non-zero  
expectation value which gives rise to the spontaneous breaking of $Z_N$ symmetry. As  
a consequence, there are $N$ degenerate states in the deconfined phase characterized 
by each element of $Z_N$.  

\smallskip

The full Euclidean action in the presence of a bosonic Higgs field $\Phi$ is given by
\begin{equation} 
S = S_G + \int_V d^3x\int_0^\beta d\tau \left[{1 \over 2}|D_\mu\Phi|^2+{m^2 \over 2}  
\Phi^\dag\Phi + {\lambda \over 4!}(\Phi^\dag\Phi)^2\right],~{\rm{\large{with}}}~D_\mu\Phi= 
\partial_\mu\Phi+ ig A_\mu\Phi. 
\end{equation} 
\noindent Here $m$ is the mass of the $\Phi$ field and $\lambda$ is the Higgs self interaction 
coupling constant. In the partition function
\begin{equation} 
{\cal Z} = \int [DA] [D\Phi] e^{-S}, 
\end{equation}  
the path-integration of $\Phi$ is over all $\Phi$ fields which are periodic in $\tau$,  
i.e $\Phi(\tau)=\Phi(\tau+\beta)$. Under the action of the above gauge transformations 
(Eq.~(\ref{bcu})), the transformed field $\Phi^\prime=U\Phi$ will not be periodic in $\tau$. 
So the actions of these gauge transformations have to be restricted to the gauge fields. 
Consequently, the action 
will increase under such gauge transformations, i.e $S(A^\prime,\Phi) > S(A,\Phi)$. It is obvious 
that the increase in the action will change if the $\Phi$ field is varied ($\Phi \to \Phi^\prime$, 
but $\Phi^\prime \ne U\Phi$ ) as the gauge fields are gauge transformed. For some $\Phi$ 
configurations, it is possible to find $\Phi^\prime$ such that 
$S(A^\prime,\Phi^\prime) = S(A,\Phi)$~\cite{Biswal:2015rul}. 
If these $\Phi$ configurations dominate the partition function, then the $Z_N$ symmetry will be 
effectively realized. In the following, we describe the simulations of the CD transition for $N=2$ 
and $m=0=\lambda$ using the above partition function.

\section{Simulations of the Confinement-Deconfinement transition} 
 
In the Monte Carlo (MC) simulations, the Euclidean space is discretized into $N_\tau\times N_s^3$ 
discrete points. $N_\tau=1/(aT)$ and $N_s=(L/a)$ are the number of lattice points along the 
temporal and spatial directions, respectively. $a$ is the lattice spacing 
and $L$ is the spatial extent of the Euclidean space. Each point $n$ on the 
lattice is represented by a set of four integers, i.e $n=(n_1,n_2,n_3,n_4)$.  
The Higgs field $\Phi_n$ 
lives on the lattice site $n$. The gauge link $U_\mu = \exp(-iagA_\mu)$, on the other hand, 
lives on the link connecting the point $n$ to its nearest neighbor along the positive 
$\mu-$direction. The action with these discretized field variables with appropriate  
scaling in terms of $a$ for $m=0=\lambda$ is given by~\cite{Kajantie:1995kf}, 
\begin{equation} 
S = \beta \sum_p {\rm{Tr}}(1-{U_p+U^\dag_p\over 2})  
- {1 \over 8} \sum_{\mu,n} {\rm{Re}}\left [(\Phi^\dag_{n+\mu}U_{n,\mu}\Phi_n)\right] 
+{1 \over 2} \sum_n\left(\Phi^\dag_n\Phi_n\right). 
\label{daction}
\end{equation} 
In Eq.~(\ref{daction}), the first term represents the pure gauge action. $U_p$ is the product 
of the gauge links going 
anti-clockwise on the $p-$th elementary square/plaquette on the lattice. 
The Polyakov loop at any spatial point $n$ is given by the path order product of  
links on the shortest temporal loop going through $n$. The gauge transformation (Eq.~(\ref{bcu})) of  
the gauge fields is equivalent to multiplication of all the temporal links on a fixed $\tau$  
slice by $z\in Z_N$. The second term represents the interaction of the gauge and Higgs 
fields. This term is not invariant under the gauge transformations (Eq.~(\ref{bcu})) of the gauge 
fields while the $\Phi$ field configuration is kept fixed. As mentioned above, the 
$\Phi$ fields can not be transformed under non-periodic gauge transformations.

\smallskip

In the Monte Carlo simulations, a sequence of statistically independent configurations of 
($\Phi_n$,$U_{\mu,n}$) are generated. This is achieved by repeatedly updating an arbitrary 
initial configuration using numerical methods which follow the Boltzmann probability factor 
$e^{-S}$ and principle of detailed balance among the configurations in the sequence. \ To update 
the gauge fields,\ we first use the standard heat bath 
algorithm~\cite{Creutz:1980zw,Kennedy:1985nu},\ and then update Higgs fields using pseudo 
heat bath algorithm~\cite{Bunk:1994xs}.\ We then again update the gauge fields using $4$ 
over-relaxation steps~\cite{Whitmer:1984he} after which Higgs fields are updated again using 
pseudo heat bath algorithm. To reduce auto-correlation between successive configurations
along the sequence (Monte Carlo history) we carry out $10$ cycles of this updating
procedure between subsequent measurements. \ For our simulations, we use the publicly 
available {\tt MILC} code~\cite{milc} and modify it to accommodate the Higgs fields.

\begin{figure}[h]
  \centering
  \subfigure[]
  {\rotatebox{270}{\includegraphics[width=0.34\hsize]
      {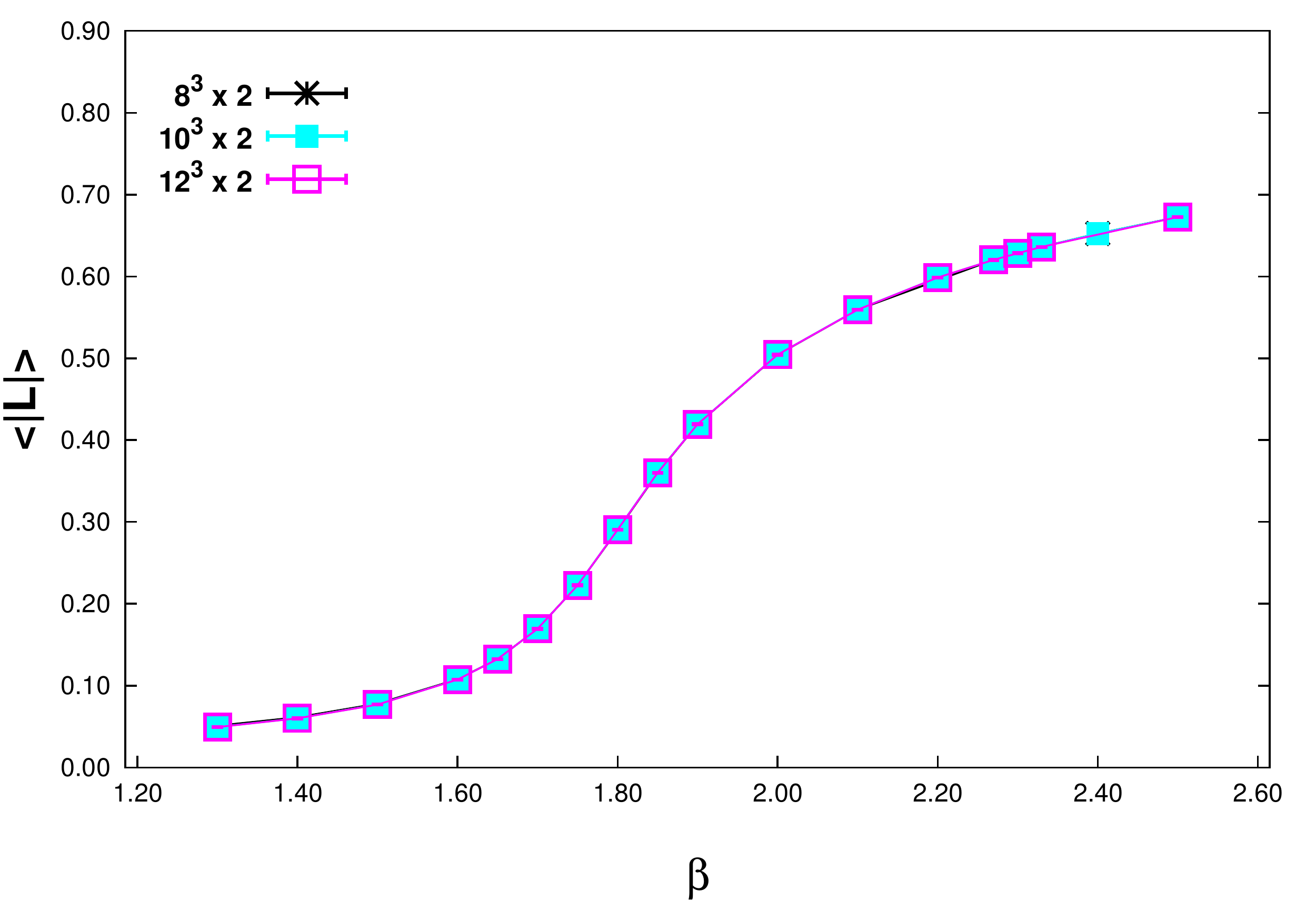}}
    \label{Fig:ploop_nt2}
  }
  \subfigure[]
  {\rotatebox{270}{\includegraphics[width=0.34\hsize]
      {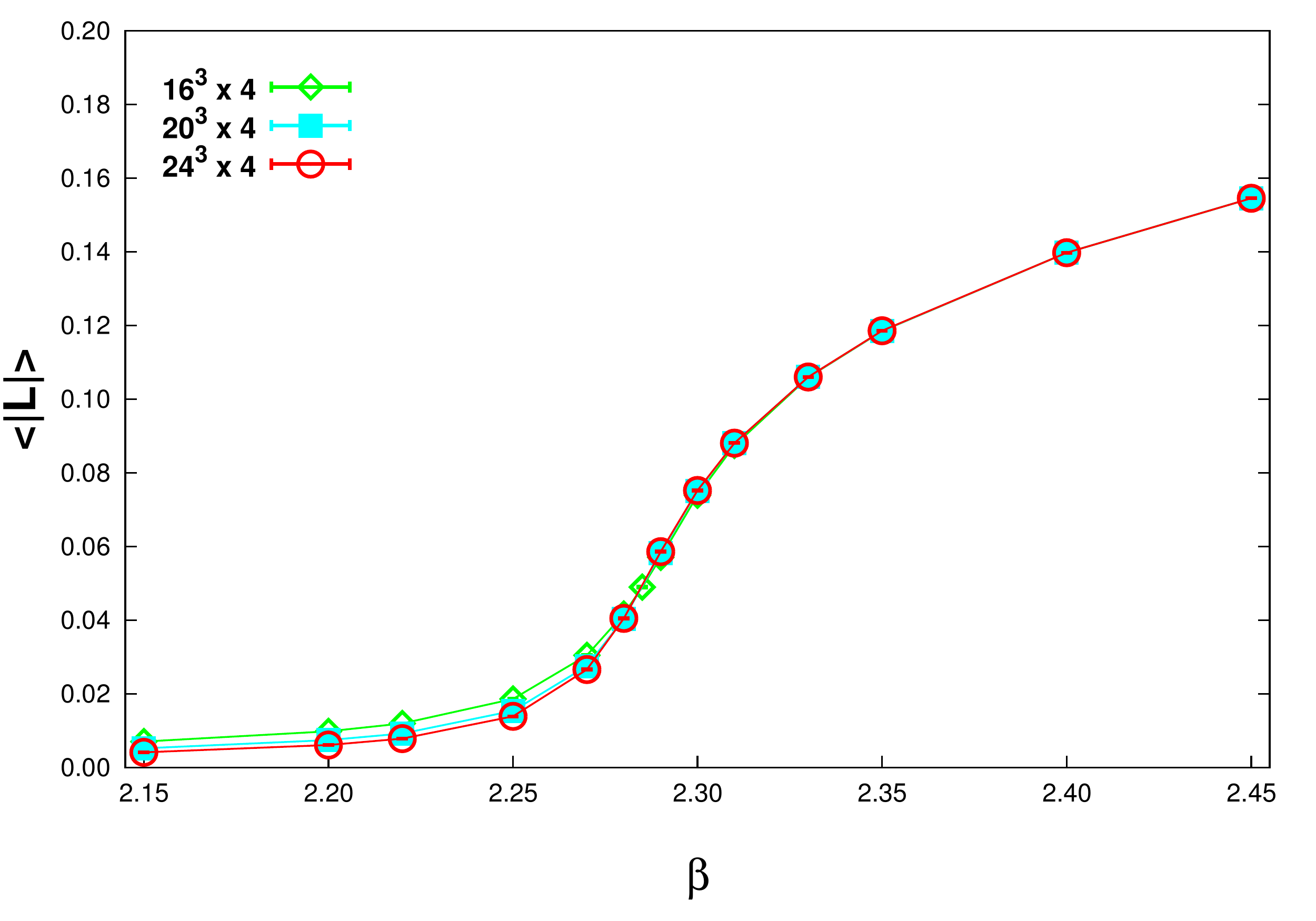}}
    \label{Fig:ploop_nt4}
  }
  \caption{The Polyakov loop average $vs$ $\beta$ for (a) $N_\tau=2$, and (b) $N_\tau=4$.}
\end{figure}

\smallskip

The CD transition is studied for three values of $N_\tau=2$, we 
consider three spatial volumes,\ $N_s = 8, 10$ and $12$. For $N_\tau=4$  we consider
$N_s=16,20$ and $24$ and for $N_\tau=8$, we 
consider $N_s = 32, 40$ and $48$. For each volume,\ we analyze $100, 000$ configurations. 
 However,  we have lower statistics for $\beta$ values far away from $\beta_c$,  
particularly for the two biggest volumes $40^3 \times 8$ and $48^3\times 8$.
The Polyakov loop, susceptibility and 
Binder cumulant are computed for various values of $\beta$ to locate 
the transition point.

 We carry out the error analysis using Jackknife method with a bin size 
of $10, 000$ configurations. We also compute the volume average of $\Phi^\dag\Phi$ and 
the interaction term. It is important to note that even though the $\Phi$ field is massless 
at the tree level, the fluctuations are finite. This is because the interaction with the gauge 
fields generate a non-zero finite mass for the $\Phi$ field. In the following section, we 
describe our simulation results.

\subsection{The CD transition for $N_\tau=2$ and $4$}

The Polyakov loop $\left<|L|\right>$ $vs$ $\beta$ for $N_\tau=2$ and
$N_\tau=4$ are shown in Figs.~\ref{Fig:ploop_nt2} and \ref{Fig:ploop_nt4}, respectively. 
$\left<|L|\right>$ grows with $\beta$ with a sharp increase around the transition. 
The $1-\rm{loop}$ $\beta-$function temperature dependence of $\left<|L|\right>$ is 
found to be consistent with the power law, 
$\left<|L|\right> \sim (T-T_c)^{1/3}$~\cite{Damgaard:1986jg}. However $\left<|L|\right>$ does not 
show any volume dependence. The peak height of the Polyakov loop susceptibility does not 
vary with volume.

\begin{figure}[h]
  \centering
  \subfigure[]
  {\rotatebox{270}{\includegraphics[width=0.34\hsize]
      {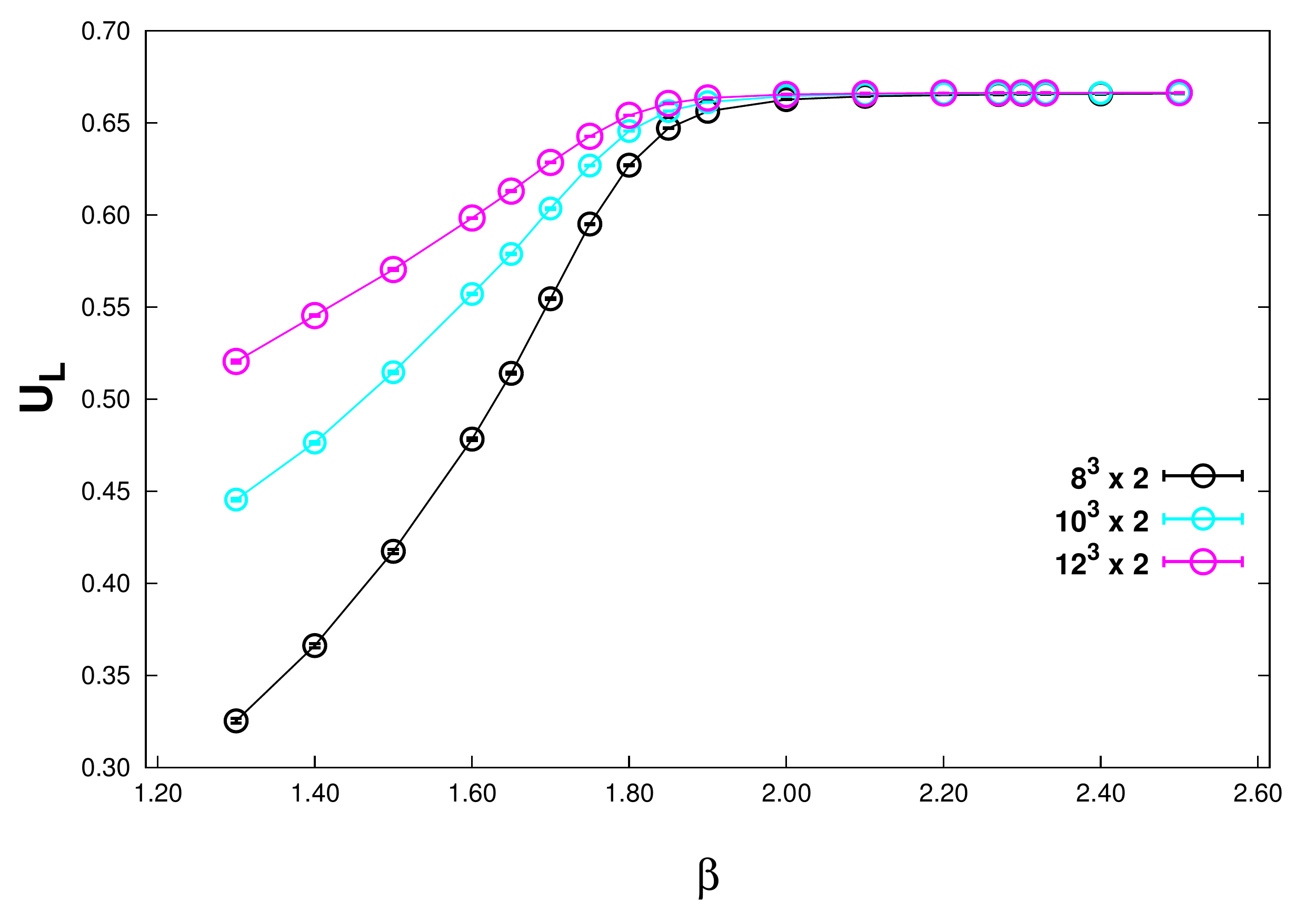}}
    \label{Fig:binder.cum_nt2}
  }
  \subfigure[]
  {\rotatebox{270}{\includegraphics[width=0.34\hsize]
      {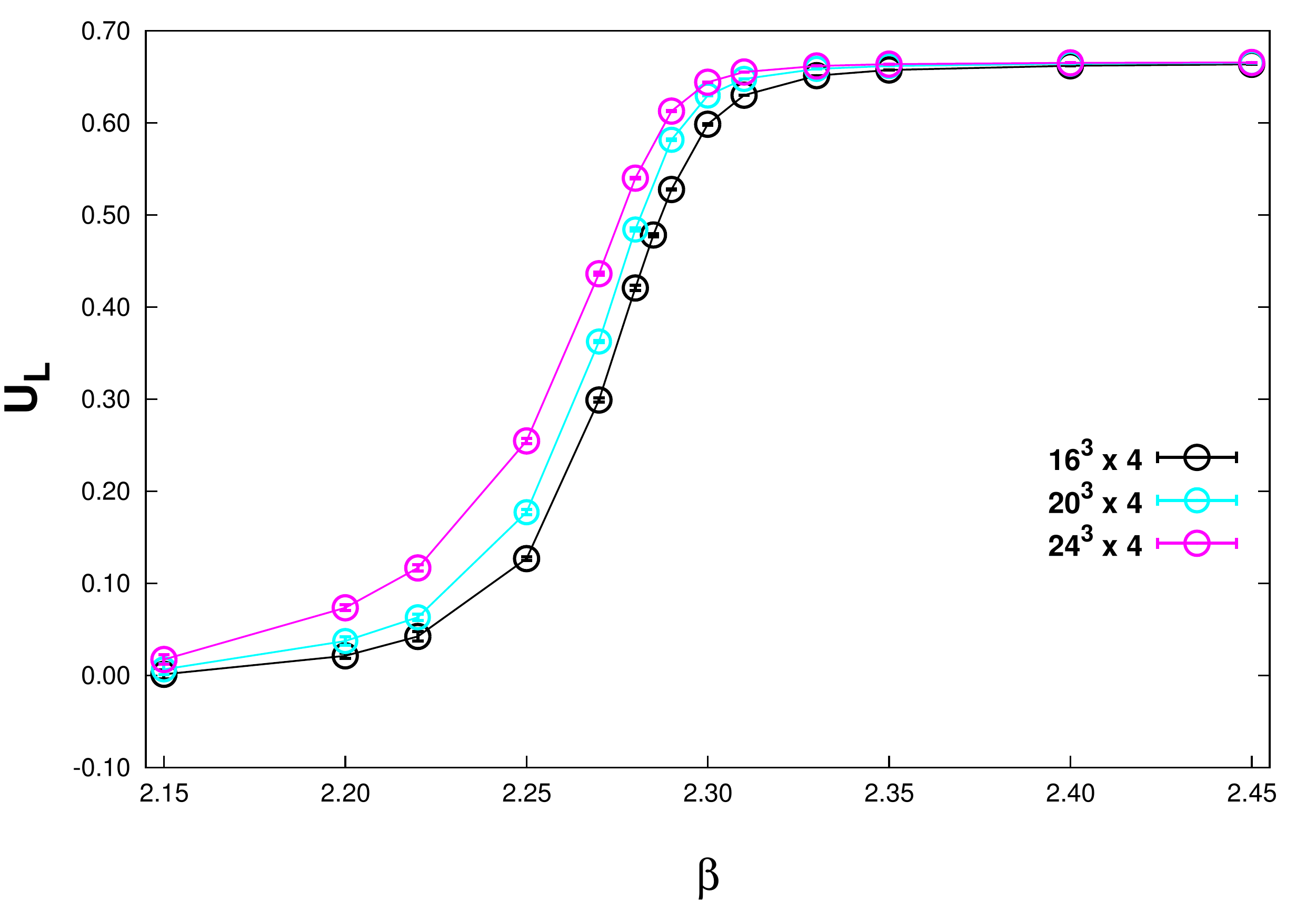}}
    \label{Fig:binder.cum_nt4}
  }
  \caption{$U_L$ $vs$ $\beta$ for different volumes for (a) $N_\tau=2$, and (b) $N_\tau=4$.}
\end{figure}

The Binder cumulant~\cite{Binder:1981sa}
\begin{equation}
U_L=1-{{\left<L^4\right>} \over {3 \left<L^2\right>^2}},
\end{equation} 
for different $\beta$ are shown in Figs.~\ref{Fig:binder.cum_nt2} and \ref{Fig:binder.cum_nt4} 
for $N_\tau=2$ and $N_\tau=4$, respectively. In both cases the variation in $U_L$ decreases 
for larger volume. For $N_\tau=2$, $U_L$ is almost flat against $\beta$. This behavior of the
Binder cumulant is exactly the opposite of what is expected in a second order phase transition. The
only explanation for these
results is that the correlation length is finite and does not grow with volume.
The sharp variation of the Polyakov loop around 
$\beta_c\sim 1.8\, (N_\tau=2)~\rm{and}~\beta_c \sim 2.29\, (N_\tau=4)$ only suggest a cross-over 
for the CD transition.

\subsection{The CD transition for $N_\tau=8$}

\begin{figure}[h]
  \centering
  \subfigure[]
  {\rotatebox{270}{\includegraphics[width=0.34\hsize]
      {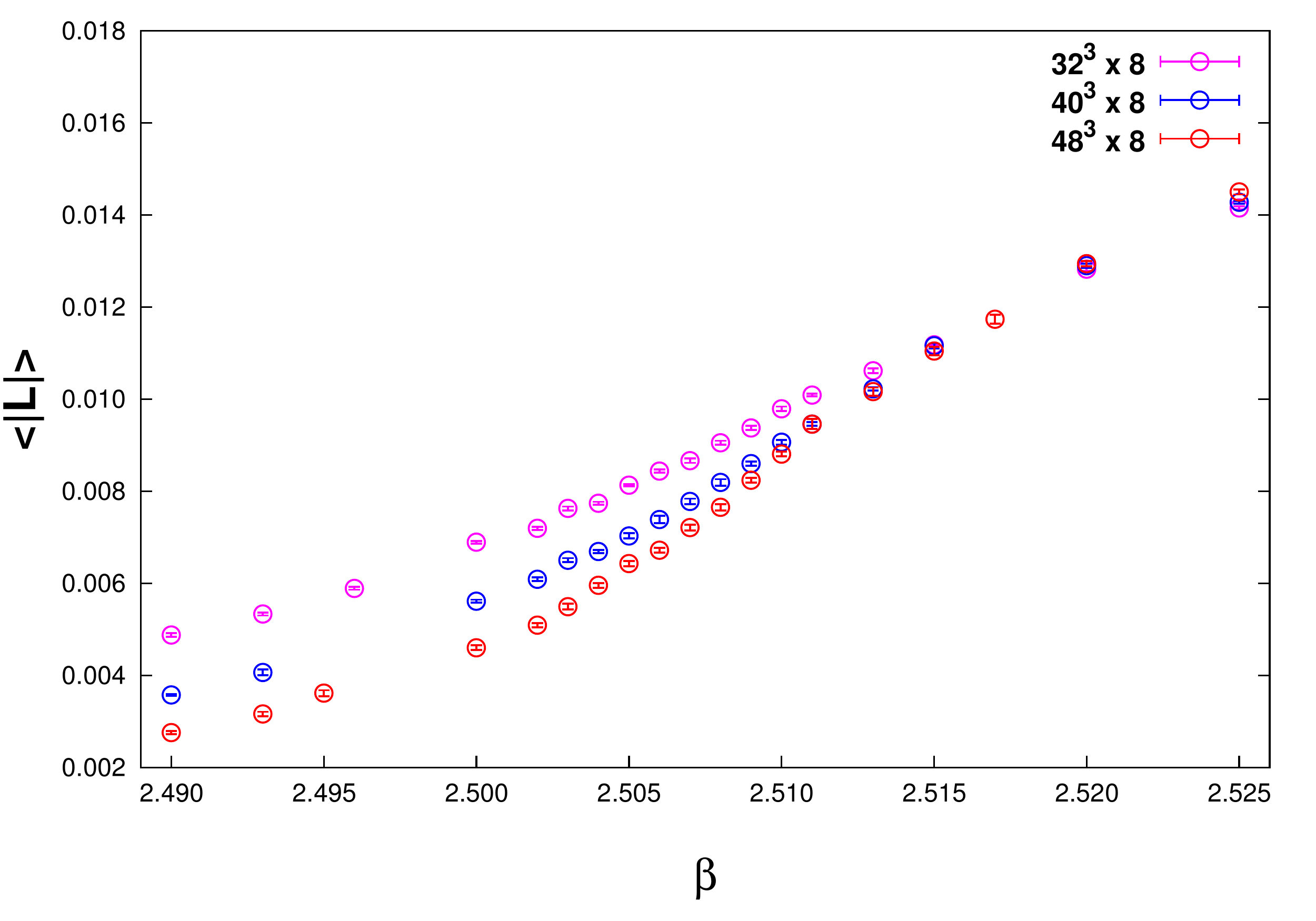}}
    \label{Fig:magtn_nt8}
  }
  \subfigure[]
  {\rotatebox{270}{\includegraphics[width=0.34\hsize]
      {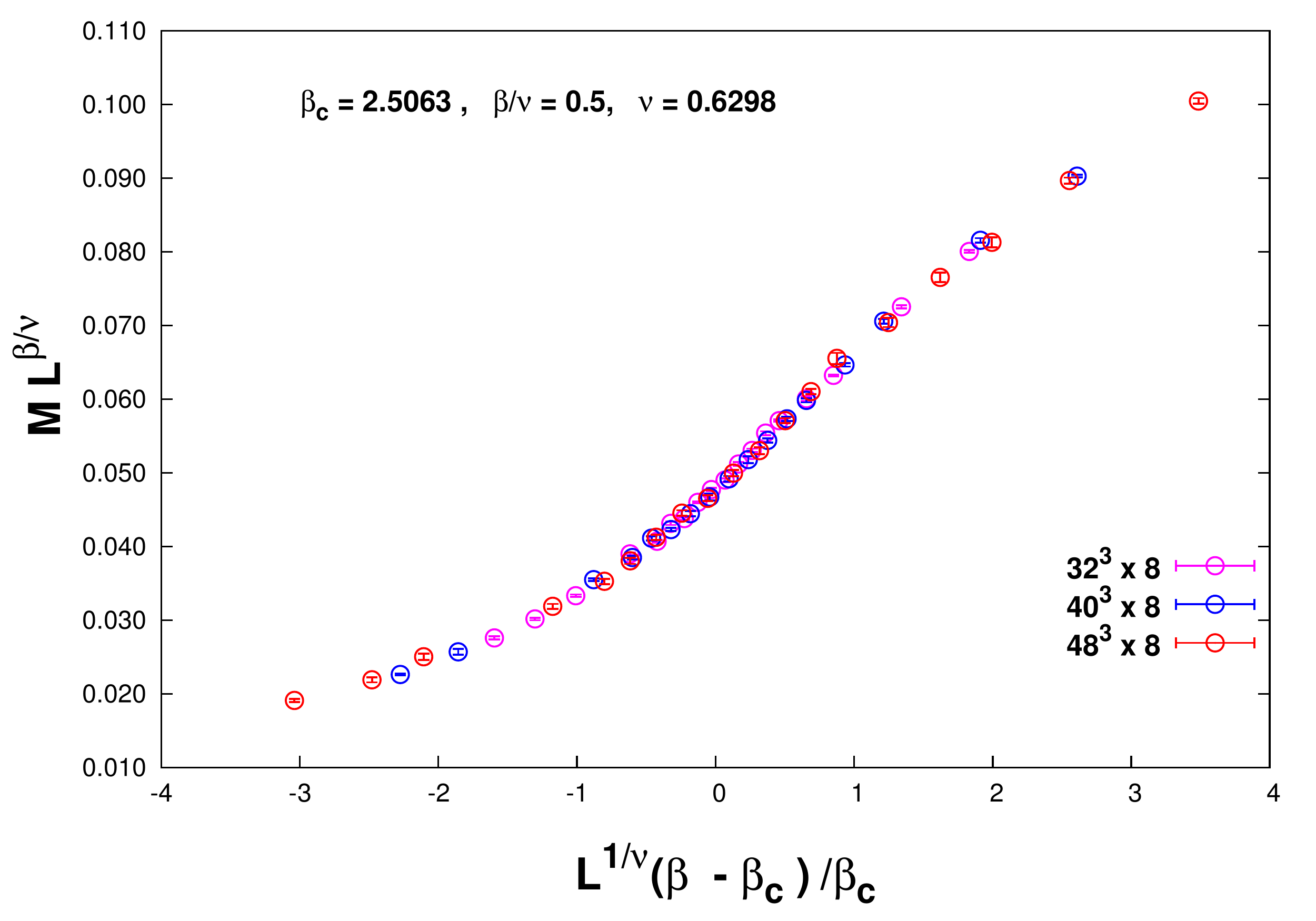}}
    \label{Fig:scaled_magtn_nt8}
  }
  \caption{$N_\tau=8$. (a) The Polyakov loop $vs$ $\beta$ for different volumes, and
    (b) Scaled Polyakov loop $vs$ $\beta$ for different volumes.}
\end{figure}
\begin{figure}[h]
  \centering
  \subfigure[]
  {\rotatebox{270}{\includegraphics[width=0.34\hsize]
      {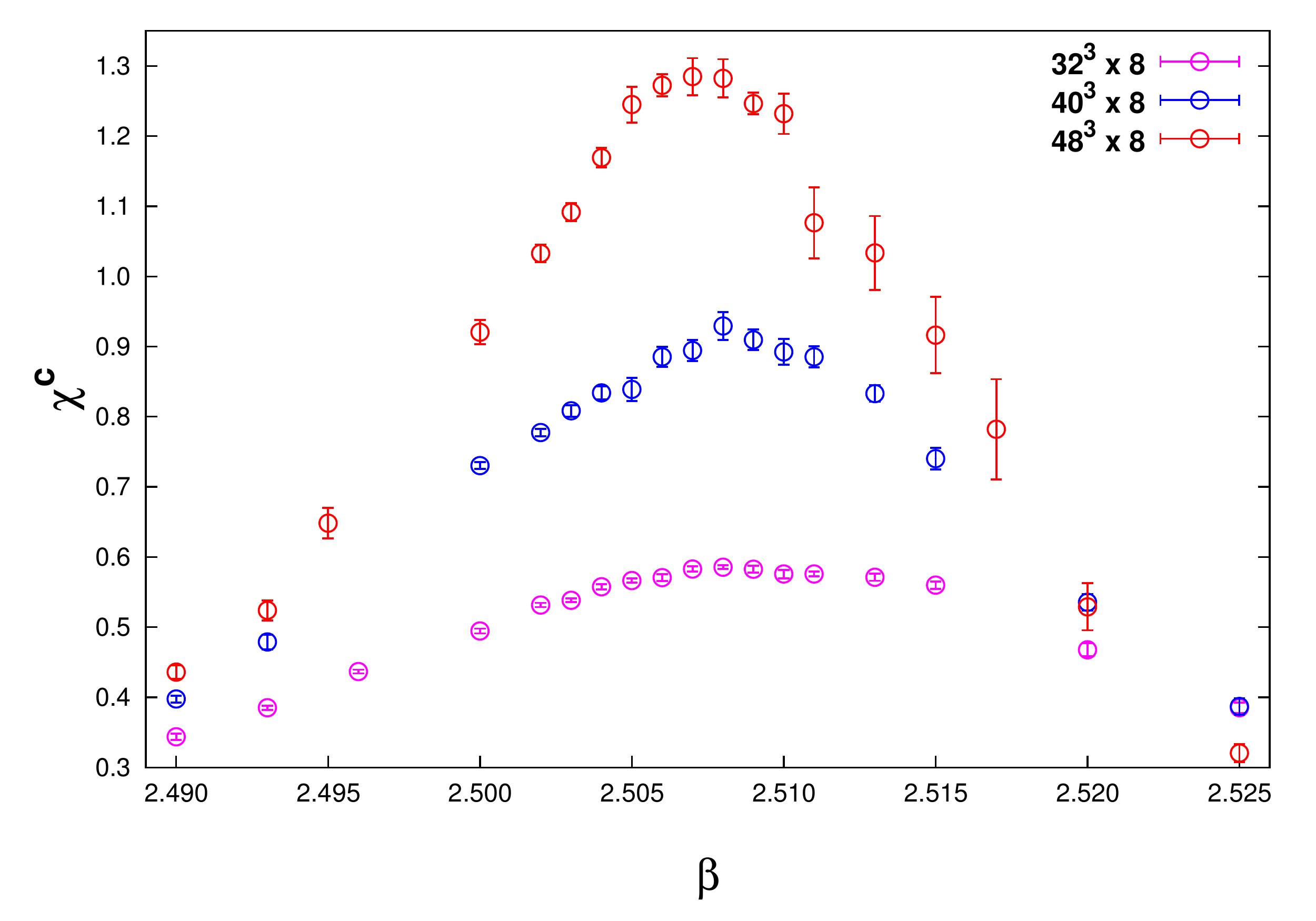}}
    \label{Fig:suspt_nt8}
  }
  \subfigure[]
  {\rotatebox{270}{\includegraphics[width=0.34\hsize]
      {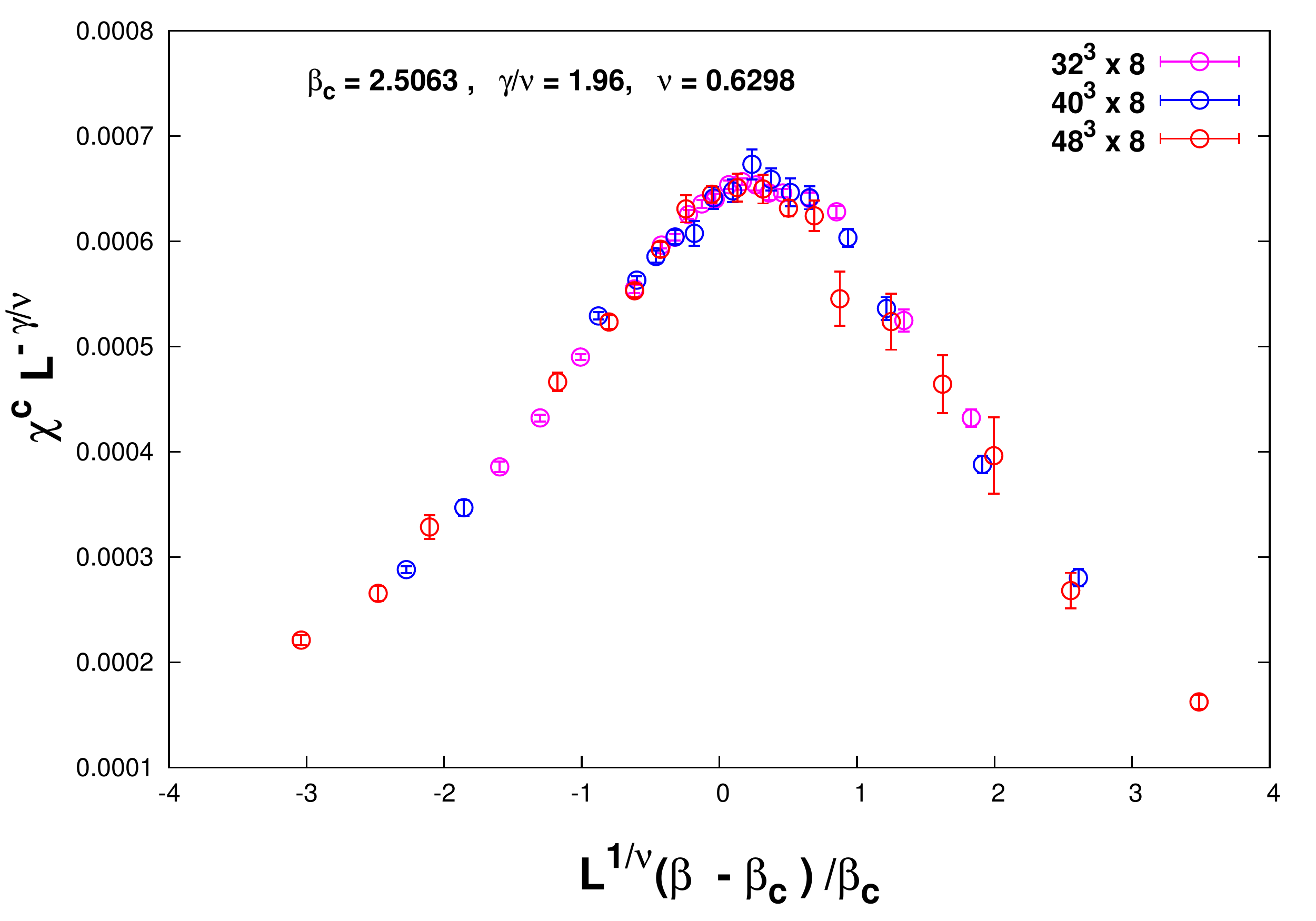}}
    \label{Fig:scaled_suspt_nt8}
  }
  \caption{$N_\tau=8$. (a) Susceptibility $vs$ $\beta$ for different volumes, and
    (b) Scaled Susceptibility  $vs$ $\beta$ for different volumes.}
\end{figure}
\begin{figure}[h]
  \centering
  \subfigure[]
  {\rotatebox{270}{\includegraphics[width=0.34\hsize]
      {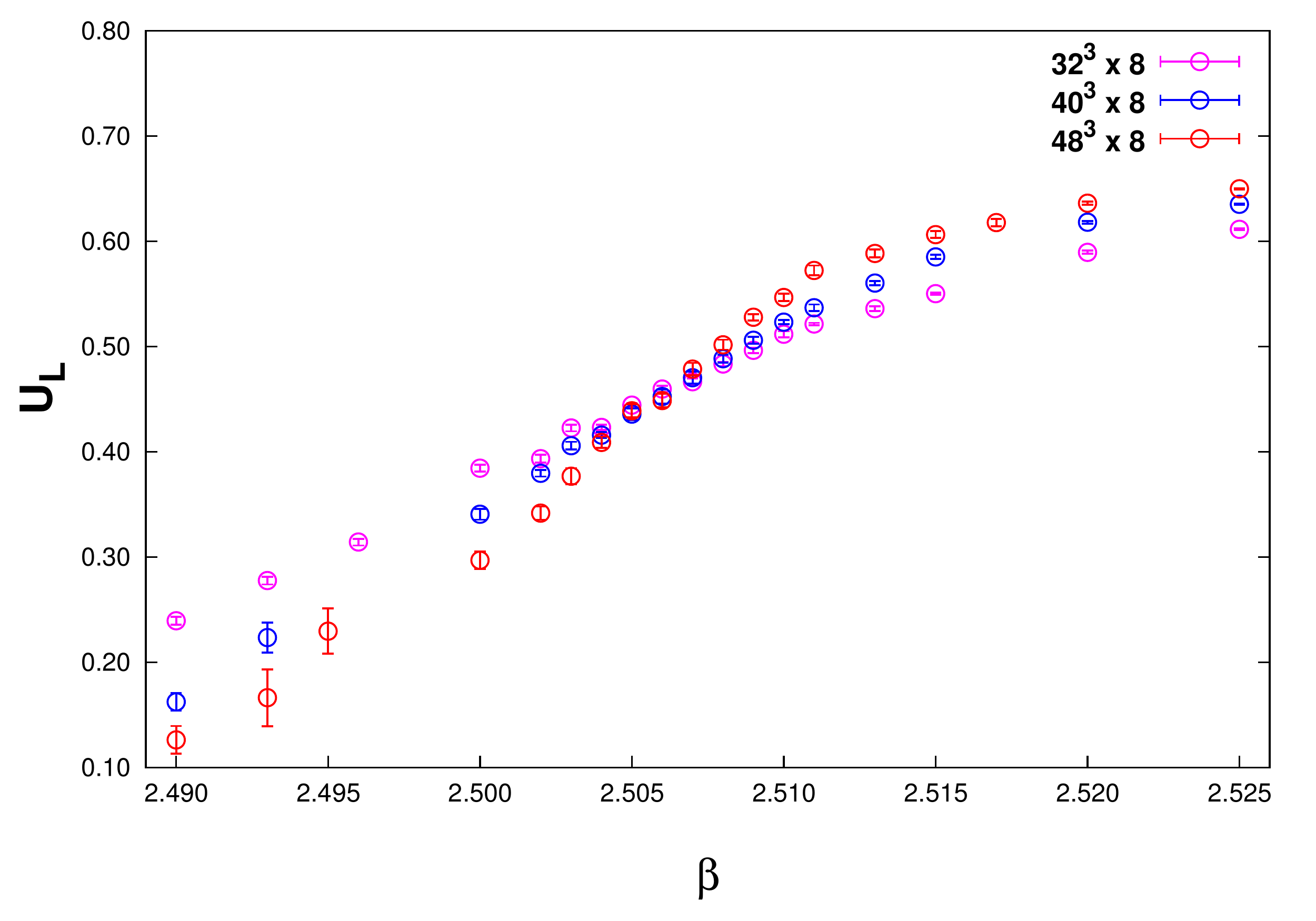}}
    \label{Fig:binder.cum_nt8}
  }
  \subfigure[]
  {\rotatebox{270}{\includegraphics[width=0.34\hsize]
      {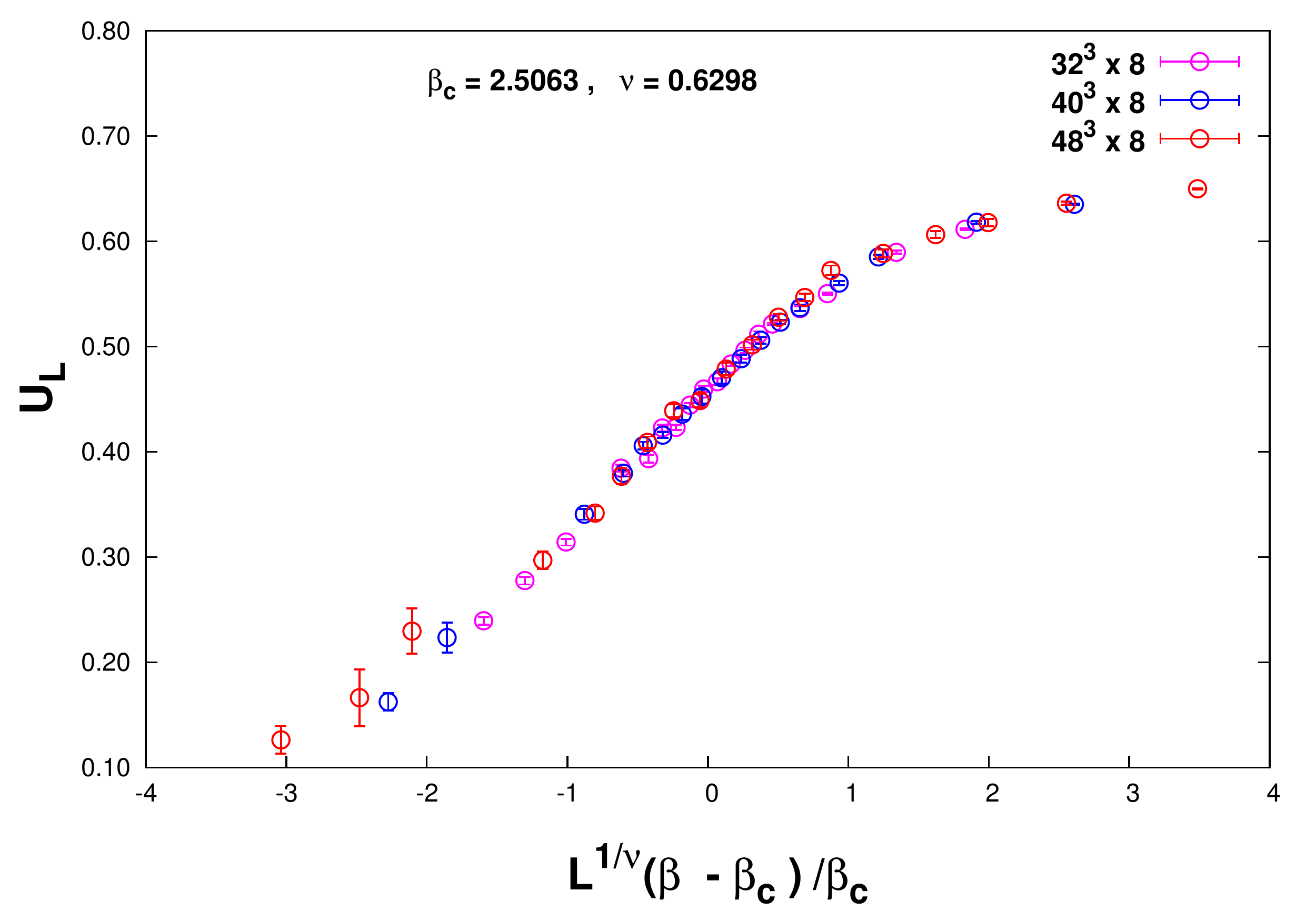}}
    \label{Fig:rescaled_binder.cum_nt8}
  }
  \caption{$N_\tau=8$. (a) The Binder cumulant $U_L$ $vs$ $\beta$ for different volumes, and
    (b) Scaled $U_L$ $vs$ $\beta$ for different volumes.}
\end{figure}
\begin{figure}[h]
  \centering
  \subfigure[]
  {\rotatebox{270}{\includegraphics[width=0.34\hsize]
      {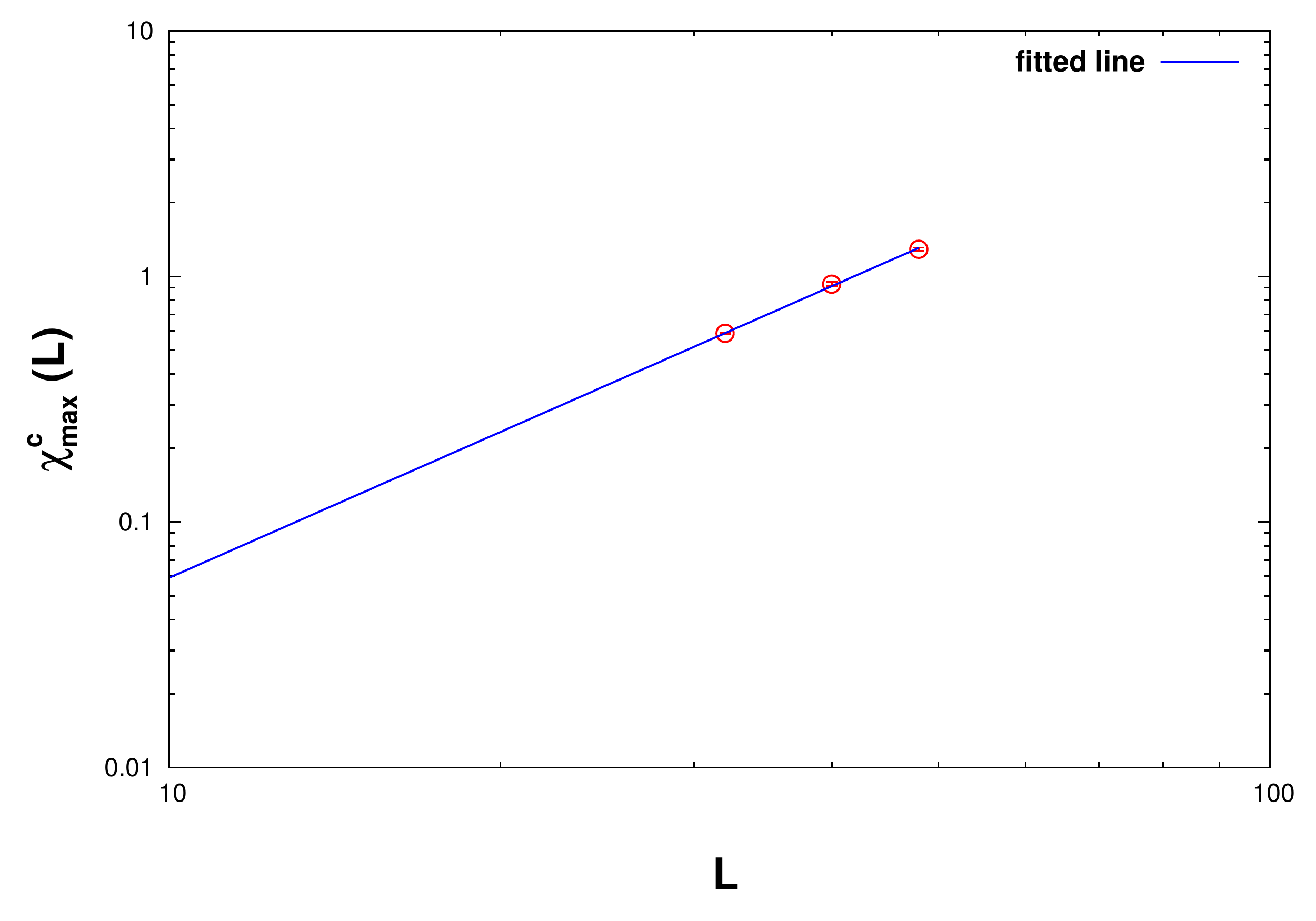}}
    \label{Fig:g2nu_succept_max_nt8}
  }
  \subfigure[]
  {\rotatebox{270}{\includegraphics[width=0.34\hsize]
      {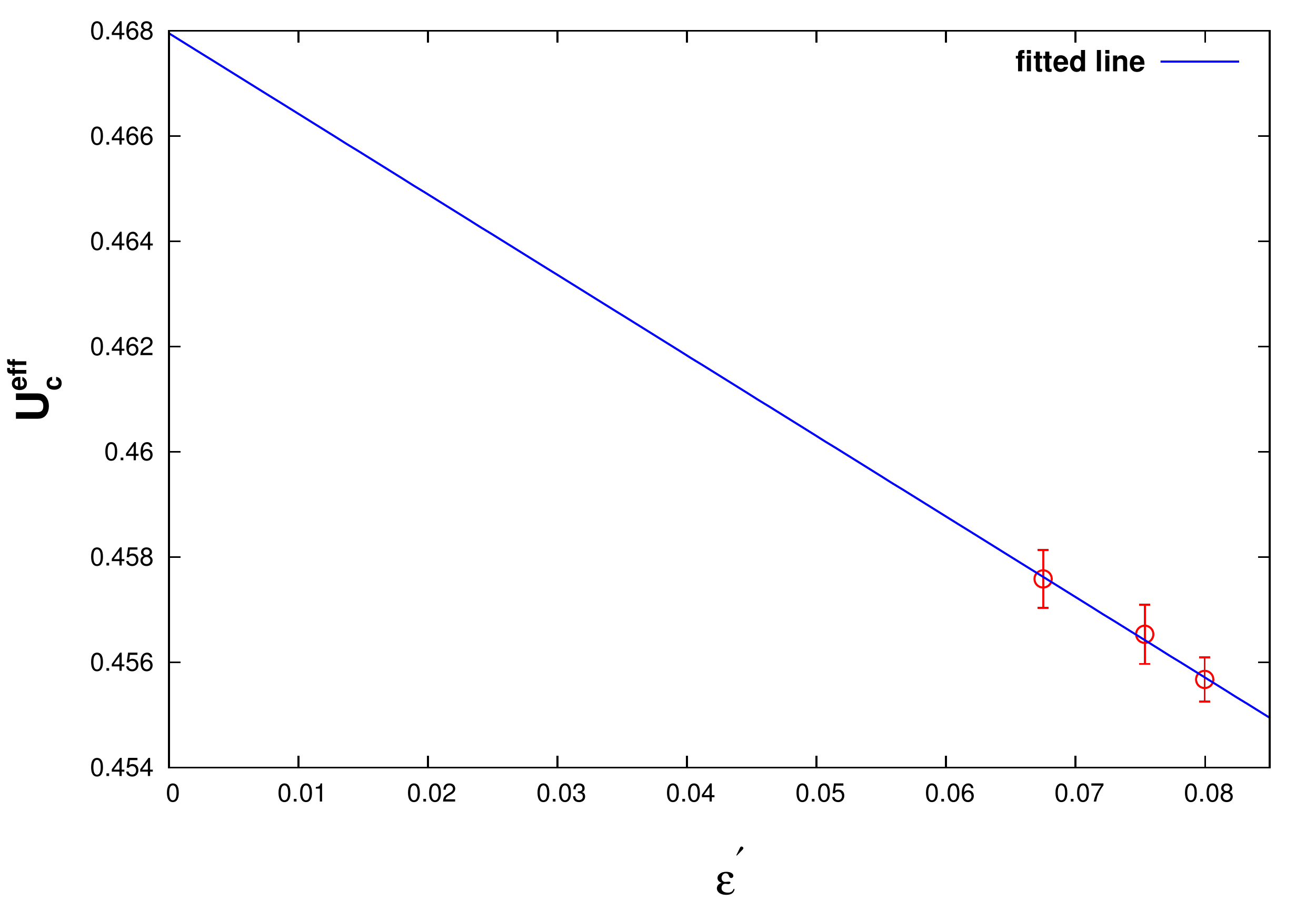}}
    \label{Fig:binder_bceff.pdf_nt8}
  }
  \caption{$N_\tau=8$. 
    (a) The values of $\chi^c_{\rmsmall{max}}$ as a function of $L$ for $L = 32, 40$ and $48$. 
    The slope of fitted line provides the value of $\gamma/\nu$. 
    (b) The values of $U_{c}^{\rm{eff}}$ obtained from the crossing points of Binder Cumulant 
    between two different volumes as a function of $\epsilon'$. The intercept provides the
    value of $U_c$.}
\end{figure}
The behavior of the Polyakov loop for $N_\tau=8$ is completely different from that of 
$N_\tau=2$ and $4$. The Polyakov loop $\left<|L|\right>$ around the transition point $\beta_c$
behaves almost like the magnetization in the Ising model. The results for 
$\left<|L|\right>$ $vs$ $\beta$ for different volumes are shown in 
Fig.~\ref{Fig:magtn_nt8}. In this case, $\left<|L|\right>$ clearly shows volume dependence. 
The volume dependence of the susceptibility $\chi^c$ of the Polyakov loop around the 
transition point is shown in Fig.~\ref{Fig:suspt_nt8}. In Figs.~\ref{Fig:scaled_magtn_nt8} 
and \ref{Fig:scaled_suspt_nt8}, we show magnetization and susceptibility 
$vs$ $(L^{1/\nu}(\beta-\beta_c)/\beta_c)$, respectively. We see that both the quantitites 
collapse to single curves.

We find the value of the exponent, $\gamma/\nu$, by studying the finite size scaling (FSS) 
of the location of the maxima of the $\chi^c$'s similar to as in~\cite{Kanaya:1994qe}. 
However instead of using Rewieghting method to determine $\chi^c_{\rmsmall{max}}$, we use 
the Cubic Spline Interpolation method to generate a few hundred points close to 
${\beta_\chi}_{\rmsmall{max}}$ for every Jackknife sample since we have reasonable amount 
of data near the peak for each volume. The scaling behavior of $\chi^c_{\rmsmall{max}}$ as 
a function of spatial volume, $L$, are shown in Fig.~\ref{Fig:g2nu_succept_max_nt8}. We 
obtain $\gamma/\nu = 1.98 (2)$.

The Binder cumulant for $N_\tau=8$ is shown in Fig.~\ref{Fig:binder.cum_nt8}.  While the 
$U_L(\beta)$ for different volumes do not intersect for $N_\tau=2$ and $4$, they do for 
$N_\tau=8$ in a narrow region around the transition point. To determine $\beta_c$ and 
corresponding value of binder cumulant, we use the following finite size behavior of $U_L$ 
in the vicintiy of the critical point,
\eqarray{
  \displaystyle U_{L} &\approx& a_0 + a_1\,  (\beta - \beta_c)/\beta_c\, L^{1/\nu} 
  + a_2\, L^{-\omega} + \cdots .
\label{eq:u_finite_dep1}
}
By following the same procedure as in~\cite{Fingberg:1992ju}, we can write 
\eqarray{
  \beta_c^{\rm{eff}} &=& \beta_c\, \left(1 - \alpha\epsilon\right)\, ,\,\,\,
  \mbox{where}\,\, \epsilon \,=\,  L^{-1/\nu - \omega} \frac{1 - b^{-\omega}}{b^{1/\nu} - 1}\,\, ,
\,\,\,\,\,\, b\, =\, \frac{L'}{L}\, ,\,\,\, b > 1 .
\label{eq:beta.c_eff_binder.cum}
}
The crossing point of the straight lines of two different spatial volumes provides 
$\beta_c^{\rm{eff}}$. By using the 3D Ising values of $\nu = 0.6298$ and $\omega = 0.825$, 
we obtain $\beta_c$ in the limit $\epsilon \rightarrow 0$ as $\beta_c = 2.5064 (4)$. 
Fig.~\ref{Fig:rescaled_binder.cum_nt8} shows that
$U_L$ $vs$ $(L^{1/\nu}(\beta-\beta_c)/\beta_c)$ for different volumes collapse to a single 
curve. To obtain infinite volume Binder Cumulant,\ $U_{c}$, we use the following relation
\eqarray{
U_{c}^{\rm{eff}} = U_{c} \left(1 + \alpha'\epsilon'\right) \, ,\,\,\,
\mbox{where}\,\, \epsilon' \,=\,  L^{- \omega} \frac{1 - b^{-\omega - 1/\nu}}{1 - b^{-1/\nu}}
\label{eq:u.c_eff_binder.cum}
}
\noindent
In Fig.~\ref{Fig:binder_bceff.pdf_nt8},\ we show $U_{c}^{\rm{eff}}$ $vs$ $\epsilon'$.\ In 
the limit $\epsilon' \rightarrow 0$,  we obtain $U_{c} = 0.468 (4)$.  To determine the 
exponent $\beta/\nu$, we find magnetization at $\beta_c$ for each volume using Cubic 
Spline Interpolation. Using $\left.\langle|L|\rangle\right|_{\beta_c} \sim L^{\beta/\nu}$, we 
get $\beta/\nu = 0.52 (2)$.

The above values of $\beta/\nu$, ${\gamma / \nu}$ and $U_L(\beta_c)$ from our
computations are close to the $3D$ Ising values. These results seem to show that the CD 
transition transition for $N_\tau=8$ is a second order phase transition.

\subsection{The $Z_2$ symmetry of the Polyakov loop}

\begin{figure}[h]
  \centering
  \subfigure[]
  {\rotatebox{270}{\includegraphics[width=0.34\hsize]
      {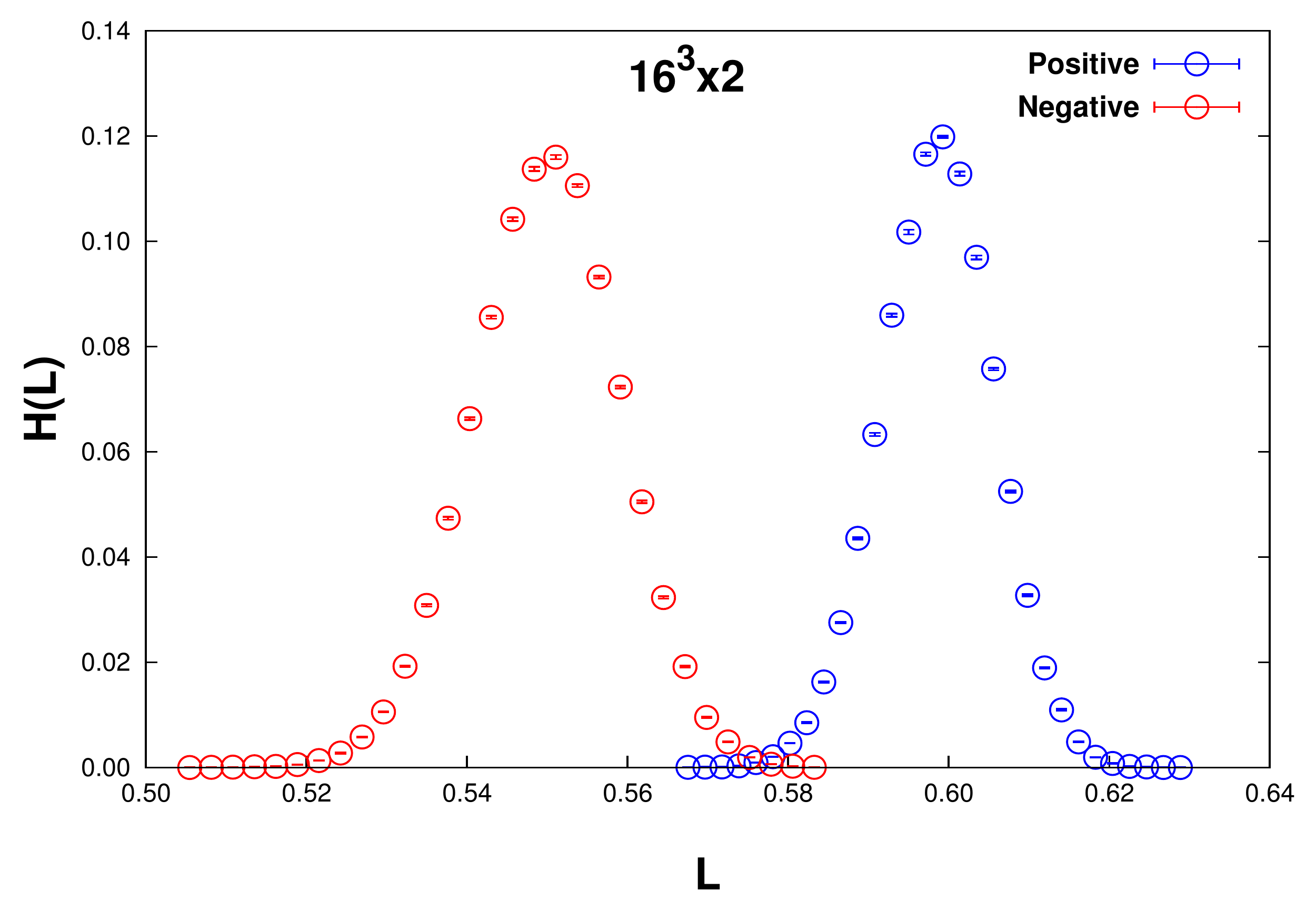}}
    \label{Fig:hl_vs_pl_nt2}
  }
  \subfigure[]
  {\rotatebox{270}{\includegraphics[width=0.34\hsize]
      {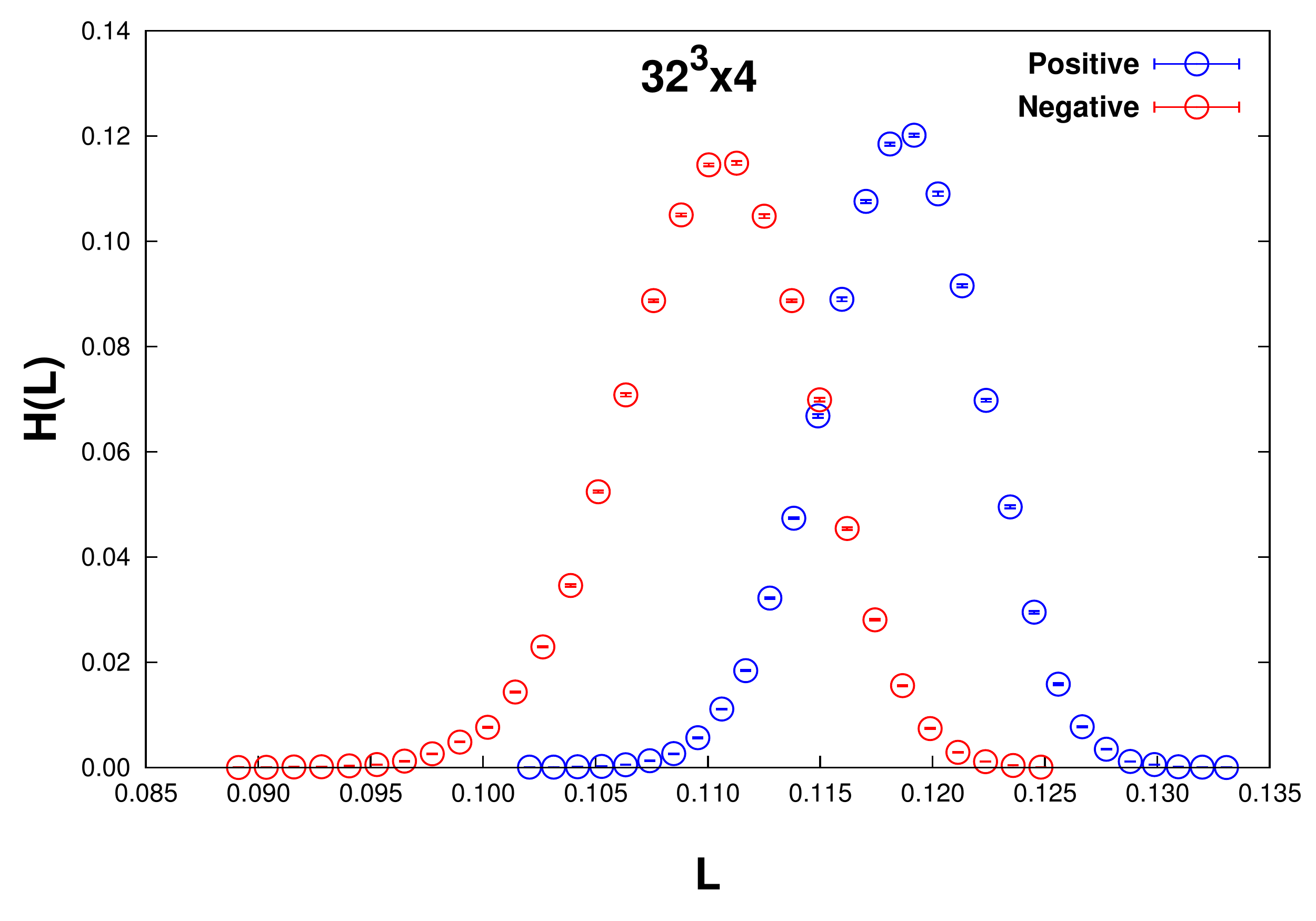}}
    \label{Fig:hl_vs_pl_nt4}
  }
  \subfigure[]
  {\rotatebox{270}{\includegraphics[width=0.34\hsize]
      {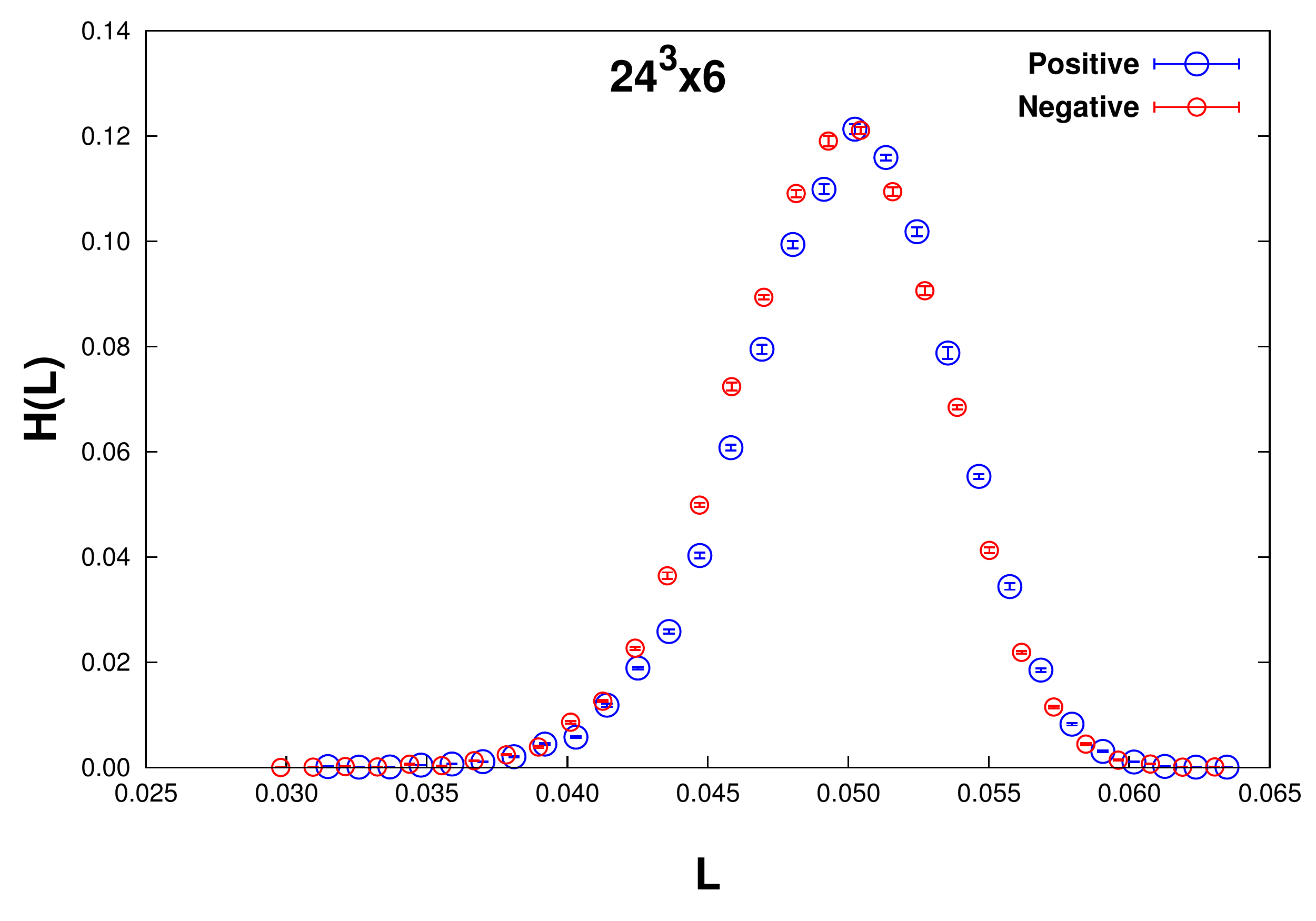}}
    \label{Fig:hl_vs_pl_nt6}
  }
  \subfigure[]
  {\rotatebox{270}{\includegraphics[width=0.34\hsize]
      {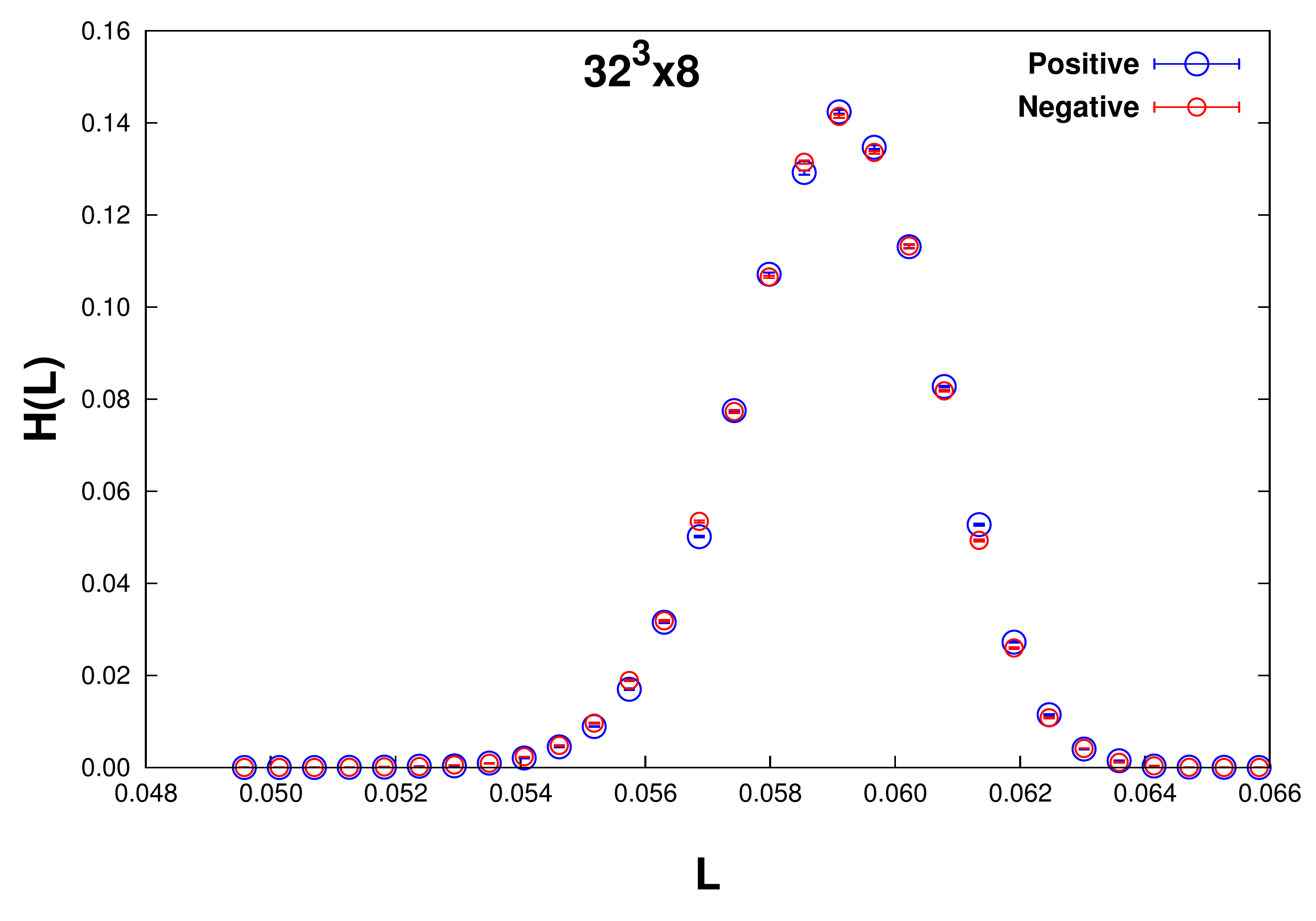}}
    \label{Fig:hl_vs_pl_nt8}
  }
  \caption{$H(L)\, vs\, |L|$. $H(L)$ is normalized to $2$. 
(a) $16^3\times 2$ lattice with $\beta=2.20$, 
(b) $32^3\times 4$ lattice with $\beta=2.35$,
(c) $24^3\times 6$ lattice with $\beta=2.50$,  
and 
(d) $32^3\times 8$ lattice with $\beta=3.20$.}
  \label{Fig:hl_vs_beta}
\end{figure}

The different $N_\tau$ studies clearly show that the nature of the CD transition 
depends on $N_\tau$. The change
in the nature of the CD transition from $N_\tau=8$ to $N_\tau=2,4$ is similar to that of 
the Ising transition when the external field is increased. So it is possible that the explicit 
breaking of the $Z_2$ symmetry 
decrease with increase in $N_\tau$. To check this, we compute the histogram of the Polyakov
loop near the transition point for $N_\tau=2,4$ and $8$. For $N_\tau=2$ and $4$, no $Z_2$ symmetry is 
observed in the distribution of the Polyakov loop. On the deconfinement side and close to
the transition point, the histograms always show one peak located on the  positive real axis.
Away from the transition point and inside the deconfinement phase, locally stable 
states are observed for which the Polyakov loop is  negative. In Fig.~\ref{Fig:hl_vs_pl_nt2} the
histogram of the Polyakov loop $H(L)$ vs $|L|$ for $\beta=2.2$ is shown for $N_\tau=2$. 
$H(L)$ is normalized to $2$.
There is no $Z_2$ symmetry either between the locations or the widths of the peaks. So
the behavior of the Polyakov loop such as thermal average,
fluctuations, correlation length etc. are found to be different for these two states. In 
contrast, the Polyakov loop exhibits $Z_2$ symmetry for $N_\tau=8$. 
Near the transition point, two peaks symmetrically located around $L=0$ on the real 
x-axis are observed. In Fig.~\ref{Fig:hl_vs_pl_nt8}, $H(L)$ $vs$ $|L|$ is
shown for $\beta=3.20$. Though $10^6$ measurements are used to compute all the data points 
in Fig.~\ref{Fig:hl_vs_pl_nt8}, each individual point in the figure is the average 
over ($H(L)*10^6$) configurations for which the Polyakov loop values belong to a small bin
centered at $L$. For example, the peaks of the histogram result from
about $\sim 1.5\times 10^4$ configurations.
It is interesting to see that $H(L)$ 
for $+L$ and $-L$ agree even with such small statistics. All physical observables which depend on 
the temporal gauge field such as gauge action and interaction term have 
same average when computed for the two $Z_2$ sector. These results suggest the 
effective realization of the $Z_2$ symmetry for $N_\tau=8$.

\begin{figure}[h]
   \centering
  {\rotatebox{270}{\includegraphics[width=0.50\hsize]
      {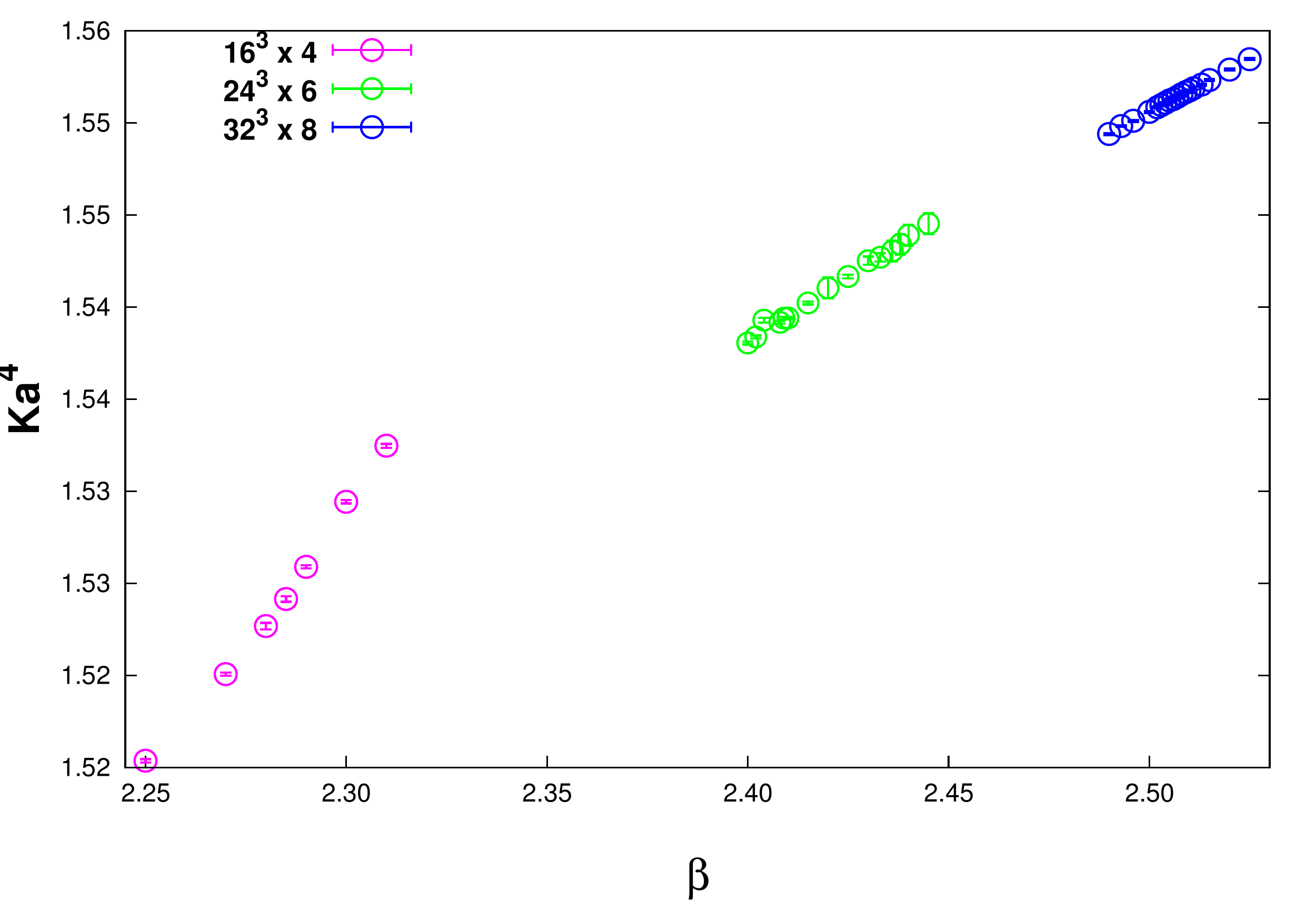}}
%    \label{Fig:int}
  }
  \caption{$Ka^4$ near $\beta_c$ for different $N_\tau$.}
\end{figure}

\section{Discussions and Conclusions}

In this work, we study the CD transition and $Z_2$ symmetry in $SU(2)+$Higgs 
theory for vanishing bare mass and quartic coupling of the Higgs field. We find that
the cut-off effects are large. For $N_\tau=2$ and $4$, the CD transition turn out to
be a crossover. The temperature dependence of the Polyakov loop average seems
to show a critical behavior above the crossover point. However, no volume dependence is
observed in any observable related to the Polyakov loop. For $N_\tau=8$, the temperature 
dependence, susceptibility and the Binder cumulant of the Polyakov loop show singular 
behavior suggesting a second order CD transition. Our results for the critical exponents 
are found to be consistent with the $3D$ Ising universality class.  

\smallskip

The singular behavior of the Polyakov loop for $N_\tau=8$ is accompanied by the 
effective realization of the $Z_2$ symmetry. $Z_2$ symmetric peaks are observed in the 
histogram of the Polyakov loop in the deconfined phase near to the transition point. 
Thermal averages such as the fluctuations of the Polyakov loop, interaction term between 
the gauge and the Higgs field, the gauge action etc. are all found to be same for the 
two deconfined states related by $Z_2$ symmetry. Note that the interaction between the 
Higgs and gauge fields are non-zero which implies that the realization of the $Z_2$ symmetry
is not due to the vanishing or small interaction.  We observe that the interaction 
in a given physical volume  increases with $N_\tau$. From $N_\tau=4$ to $6$ , the 
interaction increases by a 
factor of $\sim 5.12$ and , from $N_\tau=6$ to $8$ , it  increases by a factor 
of $\sim 3.18$.
In our simulations,  we find that
fluctuations of the Higgs field play an important role. $Z_2$ flip of the gauge fields
are always accompanied by "realignment" ($\Phi\to\Phi^\prime$) of the Higgs configuration. 
As soon as the Higgs fluctuations are frozen/fixed, the explicit breaking of $Z_2$ reappears. 
The reason why the $Z_2$ realization happens for $N_\tau=8$ and not for $N_\tau=2$ 
and $4$ is the increase in the phase space of $\Phi$ field with $N_\tau$. With the 
increase in the phase space, it is more likely that for a given $\Phi$ there exists a 
$\Phi^\prime$ which can compensate for the increase in action due to $Z_2$ rotation of the
gauge fields. We find that the likelihood of finding such a
$\Phi^\prime$ increases with $N_\tau$.
It is important to note that the $Z_2$ symmetry in our simulations only implies 
that a $\Phi^\prime$ exists for every statistically significant $\Phi$. 
It is obvious that there will be $\Phi$ configurations for which there won't
be any $\Phi^\prime$ even in the limit $N_\tau \to \infty$. This is expected 
to happen when the Higgs field acquires a condensate. In this sense, the 
restoration/realization of the $Z_2$ symmetry is not exact, and the explicit symmetry 
breaking is not zero but statistically insignificant.

\smallskip

Our results may have important implications for the study of $Z_N$ symmetry in the
presence of matter fields. Conventionally, it is expected that in the massless limit there 
will be maximal breaking of the $Z_2$ symmetry and the CD transition will be a crossover. 
$1$-loop perturbative calculations~\cite{Gross:1980br,Weiss:1981ev} for fermions suggest
that the explicit breaking for the massless case will be so large that there will be no meta-stable
states in the entire deconfinement phase.  A straightforward extension for bosonic fields
gives similar results. However, our non-perturbative results suggest
that the explicit breaking is so minimal that meta-stable states tend become degenerate
with the stable state in the continuum. It would be interesting to see if similar
realization of the $Z_N$ symmetry happens for different $N$ and also in the presence of 
fermion fields. We plan to study these issues in our future work.
\vskip 0.5cm
\centerline{\bf Acknowledgements}
All our numerical computations have been performed at Annapurna supercluster based at the Institute
of Mathematical Sciences, India and HybriLIT supercluster based at Joint Institute for Nuclear Research,
Dubna, Russia. We have used the MILC collaboration's public lattice gauge theory
code (version 6)~\cite{milc} as our base code.

%\acknowledgements

\end{document}